\documentclass{article}
\usepackage{spconf,amsmath,graphicx,hyperref}
\usepackage{cite, amsmath,amssymb, bm}
\usepackage{amsmath,amssymb, graphicx, epstopdf,cite,enumerate,booktabs, setspace, float,stfloats,bm, multirow, lscape, caption, subcaption, graphbox, soul, color, xcolor, hyperref}

\usepackage[table]{xcolor}

\newtheorem{proposition}{Proposition}

\title{
\vspace{-0.4in}
Edge Collaborative Gaussian Splatting\\with Integrated Rendering and Communication
\vspace{-0.1in}
}

\name{\begin{tabular}{c}
         Yujie Wan$^{1,2}$, Chenxuan Liu$^{2}$, Shuai Wang$^{2,*}$, Tong Zhang$^{3}$, James Jianqiao Yu$^{3}$, \\
         Kejiang Ye$^{2}$, Dusit Niyato$^{4}$, Chengzhong Xu$^{5}$
    \end{tabular}
\vspace{-0.1in}
\thanks{
This work is supported by the National Key R\&D Program of China (No. 2025YFE0204100), the Science and Technology Development Fund of Macao S.A.R (FDCT)  (No. 0074/2025/AMJ), the National Natural Science Foundation of China (Grant No. 62371444), and the Shenzhen Science and Technology Program (Grant No. RCYX20231211090206005, JCYJ20241202124934046). }
\thanks{
Corresponding author: Shuai Wang ({\tt\small s.wang@siat.ac.cn}).
}}
\address{
$^{1}$Southern University of Science and Technology \\
$^{2}$Shenzhen Institutes of Advanced Technology, Chinese Academy of Sciences \\
$^{3}$Harbin Institute of Technology, Shenzhen \ $^{4}$Nanyang Technological University
\ $^{5}$University of Macau
\vspace{-0.2in}
}
 
\begin{document}

\maketitle

\begin{abstract}
Gaussian splatting (GS) struggles with degraded rendering quality on low-cost devices.
To address this issue, we present edge collaborative GS (ECO-GS), where each user can switch between a local small GS model to guarantee timeliness and a remote large GS model to guarantee fidelity. 
However, deciding how to engage the large GS model is nontrivial, due to the interdependency between rendering requirements and resource conditions.
To this end, we propose integrated rendering and communication (IRAC), which jointly optimizes collaboration status (i.e., deciding whether to engage large GS) and edge power allocation (i.e., enabling remote rendering) under communication constraints across different users by minimizing a newly-derived GS switching function.
Despite the nonconvexity of the problem, we propose an efficient penalty majorization minimization (PMM) algorithm to obtain the critical point solution.
Furthermore, we develop an imitation learning optimization (ILO) algorithm, which reduces the computational time by over 100x compared to PMM. 
Experiments demonstrate the superiority of PMM and the real-time execution capability of ILO.
\end{abstract}

\vspace{-0.05in}

\vspace{-0.15in}
\section{Introduction}\label{sec:intro}
\vspace{-0.1in}

Gaussian splatting (GS) is an emerging paradigm for 3D reconstruction \cite{kerbl20233d}. 
However, when adapting GS to low-cost mobile devices, resource limitation necessitates model quantization or pruning, which degrades the rendering performance \cite{liu2025voyagerrealtimesplattingcityscale,gao2024cosurfgs}. 
To address the issue, a promising solution is edge collaborative GS (ECO-GS), where users can upload the pose sequences to a proximal edge server and render the corresponding images using edge GPUs \cite{Mehrabi2021MultiTier,Li2023Toward,liu2025voyagerrealtimesplattingcityscale,2025TCCN,gao2024cosurfgs}. 

The main obstacle to realizing ECO-GS is determining the collaboration timing for engaging the edge GS. 
First, collaboration may not be beneficial if the queried poses are not at the ``pain points'' (i.e., poses with significant discrepancies between rendered images and the actual environments) of the local GS model \cite{li2025Embodied}.
Second, collaboration involves periodic data exchange between user and server. 
Communication delays may lead to non-smooth rendering \cite{liu2025voyagerrealtimesplattingcityscale}. 
Existing edge collaboration schemes \cite{Mehrabi2021MultiTier,Li2023Toward,liu2025voyagerrealtimesplattingcityscale,2025TCCN,qiu2024advancingextendedreality3d,2024MIX3D,2025V2X, wang2020machine, li2025edge, zhang2024efficient}
fail in addressing the above challenges, since their objective function cannot account for the non-uniform GS discrepancies, and ignore the inter-dependency between rendering requirements and resource conditions.

To fill the gap, this paper proposes an integrated rendering and communication (IRAC) framework for ECO-GS systems, which enables users to adaptively switch between a small GS model executed locally to guarantee timeliness and a large GS model executed non-locally to enhance fidelity. 
Specifically, we propose a novel GS switching model to distinguish the heterogeneous view discrepancies of different GS models. 
The new model serves as objective function and is integrated with the communication resource constraints for cross-layer IRAC optimization, enabling us to maximize the information gain brought by collaboration. 
Next, the IRAC problem involves both binary edge-device collaboration and continuous power control variables.
Existing methods \cite{diamond2016cvxpy,yu2004iterative,wang2020machine,sun2016majorization,zhang2024efficient,hubner2014rounding} may lead to ineffective solutions, since they involve continuous relaxation of discrete variables. 
To tackle the challenge, we propose penalty majorization minimization (PMM) and imitation learning optimization (ILO) methods to solve the problem for high-quality real-time IRAC designs.
Finally, experimental results demonstrate the superiority of IRAC-based ECO-GS over benchmark methods in various scenarios. 


\vspace{-0.15in}
\section{Problem Formulation}\label{sec:prob}
\vspace{-0.1in}

\begin{figure}[!t]
\centering 
\includegraphics[width=0.98\linewidth]{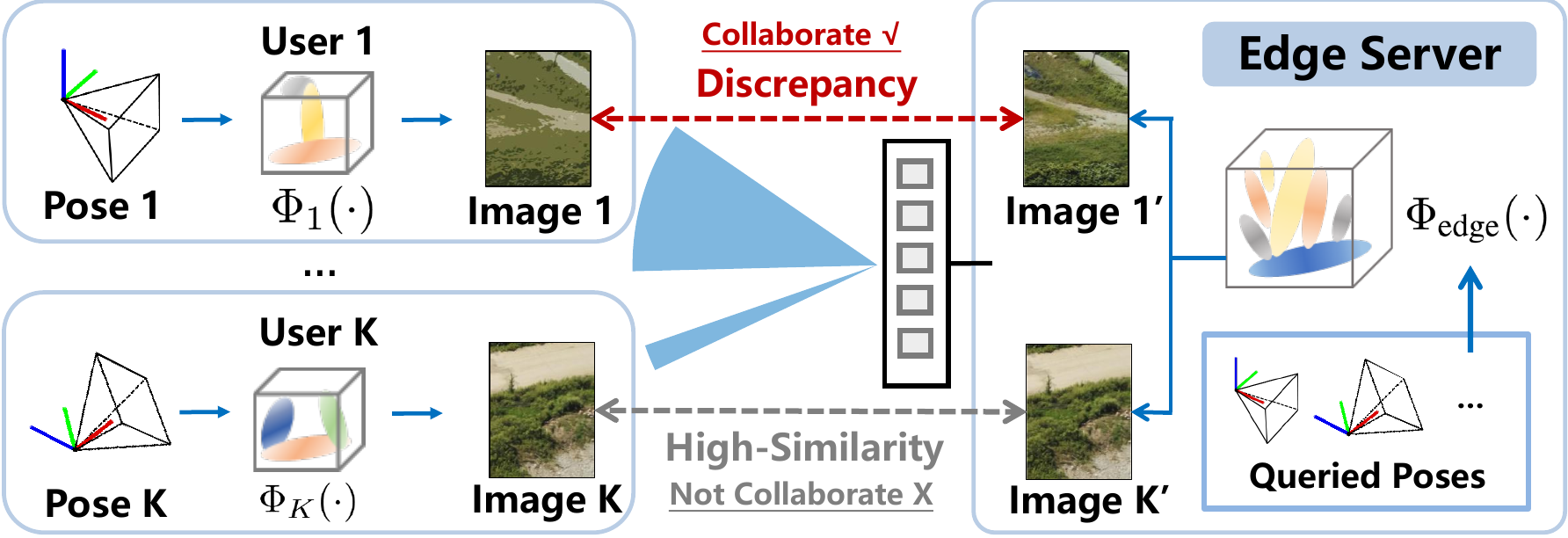}
\vspace{-0.05in}
\caption{The ECO-GS system.}
\label{fig1}
\vspace{-0.15in}
\end{figure}

We consider an ECO-GS system shown in Fig.~\ref{fig1}, which consists of an edge server with $N$ antennas and $K$ single-antenna users ($N\gg K$).
The ECO-GS system aims to render RGB images $\{\mathbf{v}_{k}\in \mathbb{R}^{3LW}\}$ at all $K$ users given their 6D view poses $\{\mathbf{s}_{k}\in \mathbb{R}^{6}\}$, where $L$ and $W$ are the length and width of camera images, respectively \cite{2025TCCN}. 
For each $\mathbf{v}_{k}$, it can be either rendered by a local GS model $\Phi_{k}(\cdot)$, or a remote edge GS model $\Phi_{\mathrm{edge}}(\cdot)$.
A binary vector $\mathbf{x}=[x_1,\cdots, x_K]^T$ with $x_k\in\{0,1\}$ is introduced, where $x_k=1$ indicates that the $k$-th user adopts edge GS, and $x_k=0$ otherwise. 
The output of ECO-GS is thus
\begin{align}\label{moe}
&\mathbf{v}_{k}= (1-x_k)\,\Phi_{k}(\mathbf{s}_{k}) + x_k\,\Phi_{\mathrm{edge}}(\mathbf{s}_{k}).
\end{align}

For a user with $x_k=1$, it needs to download $\mathbf{v}_{k}$ from the server, where the image data volume (in bits) is $V_k$.
This procedure can be modeled as $r_k = \sum_{j=1}^K x_j\mathbf{h}_{k}^H \mathbf{w}_jz_{j}+n_k$,
where $r_k\in\mathbb{C}$ is the received signal, $\mathbf{h}_{k} \in \mathbb{C}^{N \times 1}$ is the downlink channel from the server to the $k$-th user, $\mathbf{w}_k\in\mathbb{C}^{N \times 1}$ is the beamforming vector with power $\|\mathbf{w}_k\|^2=p_k$, $z_{k}\in\mathbb{C}$ is the modulated signal with power $\mathbb{E}[|z_{k}|^2]=1$, and $n_k\in\mathbb{C}$ is additive white Gaussian noise (AWGN) with zero mean and variance $\sigma_k^2$. 
To decode $z_k$, we employ the maximum ratio combining (MRC) beamformer $\mathbf{w}_k=
\sqrt{p_k}\left\Vert\mathbf{h}_{k}\right\Vert_2^{-1}\mathbf{h}_{k}$, which is asymptotically optimal when $N\gg K$ \cite{song2025distributed}.
The data rate of user $k$ is given by
$R_{k}(p_k)=B_k\mathrm{log}_2\left(1+\Vert\mathbf{h}_{k}\Vert_2^2p_{k}/\sigma^2_k\right)$,
where $B_k$ in Hz is the bandwidth of user $k$ and we have adopted 
$|\mathbf{h}_k^H\mathbf{h}_{j}|^2/\left\Vert\mathbf{h}_{k}\right\Vert_2^2 \rightarrow 0$ for $k\neq j$ as $N\rightarrow+\infty$ \cite{song2025distributed,wang2020angle}.

In ECO-GS systems, the design variables are the collaboration status vector $\mathbf{x}=[x_1,\cdots,x_K]^T$ and transmit power vector $\mathbf{p}=[p_1,\cdots,p_K]^T$.
The edge transmit power should not exceed its budget $P$, i.e., $\sum_{k=1}^Kp_k\leq P$ \cite{yu2004iterative, zhang2024efficient}. 
To ensure smoothness, remote rendering should be completed within a time duration $T$, i.e., $T_0+x_kV_kR_k^{-1}\leq T$ \cite{liu2025voyagerrealtimesplattingcityscale}, where 
$T_0$ is edge rendering time and $x_kV_kR_k^{-1}$ is image downloading time. 
To avoid computation overload, the server supports at most $S$ users, i.e., $\sum_{k=1}^Kx_k\leq S$ \cite{li2025edge}.
Having the resource constraints satisfied, it is then crucial to minimize the discrepancy between 
the rendered image $\mathbf{v}_{k}$ and the ground truth image $\widehat{\mathbf{v}}_{k}$.
To quantify the deviation from the desired image, we adopt a \emph{rendering error function} based on \cite{kerbl20233d}, defined as:
\begin{equation}
\mathcal{L}\left(\mathbf{v}_{k}, \widehat{\mathbf{v}}_{k}\right)
= (1-\lambda)\|\mathbf{v}_{k}-\widehat{\mathbf{v}}_{k}\|_1
+ \lambda \mathcal{L}_{\mathsf{DS}}\left(\mathbf{v}_{k},\widehat{\mathbf{v}}_{k}\right),
\label{eq:gs_loss}
\end{equation}
where the structural dissimilarity metric is computed as 
$
\mathcal{L}_{\mathsf{DS}}\left(\mathbf{v}_{k}, \widehat{\mathbf{v}}_{k}\right) = 1 - \text{SSIM}\left(\mathbf{v}_{k}, \widehat{\mathbf{v}}_{k}\right)$,
with $\text{SSIM}$ being the structural similarity index measure function detailed in \cite[Eqn. 5]{wang2011ssim}, and 
$\lambda$ is a small weighting factor \cite{kerbl20233d}.
Combining the above leads to the \textbf{IRAC} problem below:
\begin{subequations}
\setlength{\abovedisplayskip}{1pt}
\begin{align}
	   \! \!\! \! 	\text{P}: \min_{\mathbf{x},\mathbf{p}}~~&\sum_{k=1}^K  \mathcal{L}\left(\mathbf{v}_{k}, \widehat{\mathbf{v}}_{k}\right) \label{Pa}  \\
		 \! \!\! \! 	\textrm{s.t.} ~~ & 
 B_k\mathrm{log}_2\left(1+\frac{\Vert\mathbf{h}_{k}\Vert_2^2p_{k}}{\sigma^2_k}\right)\geq \frac{x_kV_k}{T-T_0}, \ \forall k,  \label{Pb} \\
   \! \! \! \!  & \sum_{k=1}^{K}p_{k} \leq P, \ \sum_{k=1}^K x_{k} \le S, \label{Pc} \\
      \! \! \! \!  & p_k\geq 0, \ x_k\in\{0,1\}, \ \forall k. \label{Pd}
\end{align}
\end{subequations}

Solving problem P is challenging due to: (i) the inaccessibility of ground truth images, i.e., the cost function $\mathcal{L}\left(\mathbf{v}_{k}, \widehat{\mathbf{v}}_{k}\right)$ cannot be evaluated directly; (ii) P involves coupling between binary variables $\mathbf{x}$ and continuous variables $\mathbf{p}$.

\vspace{-0.1in}
\section{Penalty Majorization Minimization}\label{sec:pmm}
\vspace{-0.1in}

To tackle challenge (i), we find a surrogate cost function of problem P.
According to the triangle inequality, we have 
\begin{align}\label{eq:triangle}
&\|\widehat{\mathbf{v}}_{k} -\Phi_{k}(\mathbf{s}_{k})\|_1
-
\|\Phi_{\mathrm{edge}}(\mathbf{s}_{k})-\widehat{\mathbf{v}}_{k}\|_1
\nonumber\\
&\leq
\|\Phi_{\mathrm{edge}}(\mathbf{s}_{k})-\Phi_{k}(\mathbf{s}_{k})\|_1.
\end{align}
Putting \eqref{eq:triangle} into \eqref{eq:gs_loss}, and with a sufficiently small $\lambda$, we have 
$$
\mathcal{L}\left(\widehat{\mathbf{v}}_{k},\Phi_{k}(\mathbf{s}_{k})\right)-
\mathcal{L}\left(\Phi_{\mathrm{edge}}(\mathbf{s}_{k}),\widehat{\mathbf{v}}_{k}\right)
\leq
\mathcal{L}\left(\Phi_\mathrm{edge}(\mathbf{s}_{k}),\Phi_{k}(\mathbf{s}_{k})\right).
$$
Putting the above inequality and \eqref{moe} into \eqref{eq:gs_loss}, we have 
\begin{align}\label{eq:ub}
&\mathcal{L}\left(\mathbf{v}_{k}, \widehat{\mathbf{v}}_{k}\right)=
(1-x_k)\mathcal{L}\left(\Phi_{k}(\mathbf{s}_{k}), \widehat{\mathbf{v}}_{k}\right)
+x_k\mathcal{L}\left(\Phi_{\mathrm{edge}}(\mathbf{s}_{k}), \widehat{\mathbf{v}}_{k}\right)
\nonumber\\
&
\leq \underbrace{\mathcal{L}\left(\Phi_\mathrm{edge}(\mathbf{s}_{k}), \widehat{\mathbf{v}}_{k}\right)}_{\text{const.}}
+
\underbrace{(1-x_k)\mathcal{L}\left(\Phi_\mathrm{edge}(\mathbf{s}_{k}),\Phi_{k}(\mathbf{s}_{k})\right)}_{\text{var.}}.
\nonumber
\end{align}
As such, minimizing $\mathcal{L}$ in problem P can be safely converted to minimizing its upper bound, which is now an explicit function of $x_k$.
Problem P thus becomes 
  \begin{align}
    \text{P}1:\min_{\substack{\mathbf{x},\mathbf{p}}} 
      \ &\sum_{k=1}^K{(1-x_k)\mathcal{L}\left(\Phi_\mathrm{edge}(\mathbf{s}_{k}),\Phi_{k}(\mathbf{s}_{k})\right)} + \text{const.}
      \nonumber \\
    \ \text{s.t.} \ \
     &\eqref{Pb}, \ \eqref{Pc}, \ \eqref{Pd},
  \end{align} 
where term $\text{const.}$ can be ignored.
The new objective function in P1 is termed \textbf{GS switching} function, which quantifies the difference between collaborative and non-collaborative rendering. 
This function no longer involves ground truth images, and can be computed using GS models $\Phi_\mathrm{edge}$ and $\{\Phi_k\}$. 
Consequently, we can deploy all GS models at the server to evaluate the GS switching cost.

To tackle challenge (ii), we relax the binary constraint $x_{k}\in\{0,1\}$ in \eqref{Pd} into an affine constraint $x_{k}\in [0,1]$. 
To promote a binary solution for the relaxed variable $\{x_{k}\}$, we augment the objective function with a penalty term \cite{lucidi2010exact}: $\varphi(\mathbf x)=\frac{1}{\beta}\sum_{k=1}^K x_{k}(1-x_k)$,
where $\beta>0$ is the penalty parameter.
Then problem P1 is transformed into 
  \begin{align}
    \text{P}2:\min_{\substack{\mathbf{x},\mathbf{p}}} 
      \quad  &\sum_{k=1}^K{(1-x_k)\mathcal{L}\left(\Phi_\mathrm{edge}(\mathbf{s}_{k}),\Phi_{k}(\mathbf{s}_{k})\right)}
    +
\varphi(\mathbf x)
      \nonumber \\
    \ \text{s.t.} \ \
     &p_k\geq0, \ 0\leq x_k \leq 1, \ \forall k, \ \eqref{Pb}, \ \eqref{Pc}.
  \end{align} 
According to \cite[Proposition 1]{lucidi2010exact}, there always exists a $\beta$ such that P1 and P2 are equivalent problems.

To further tackle the non-convexity of $\varphi(\mathbf x) $, we propose to leverage the MM framework, which solves a sequence of surrogate problems $(\text{P}2[1],\text{P}2[2],\cdots)$ of P2 in an iterative fashion. 
Specifically, given solution $\mathbf{x}{[n]}$ at the $n$-th iteration, problem $\text{P}2[n+1]$ is considered at the $(n+1)$-th iteration:
  \begin{align}
    \min_{\substack{\mathbf{x},\mathbf{p}}} 
      \  &\sum_{k=1}^K{(1-x_k)\mathcal{L}\left(\Phi_\mathrm{edge}(\mathbf{s}_{k}),\Phi_{k}(\mathbf{s}_{k})\right)}
    +
\widehat{\varphi}(\mathbf x|\mathbf x^{[n]})
      \nonumber \\
    \ \text{s.t.} \ \
     &p_k\geq0, \ 0\leq x_k \leq 1, \ \forall k, \ \eqref{Pb}, \ \eqref{Pc}.
  \end{align} 
where the surrogate function is given as follows:
\begin{align}
\widehat{\varphi}(\mathbf x|\mathbf x^{[n]}) &= 
\sum_{k=1}^K 
\left(
\frac{1}{\beta}x_{k}-\frac{2}{\beta}x_{k}^{[n]}x_{k}+\frac{1}{\beta}x_{k}^{[n]^2}
\right).
\end{align}

\begin{proposition}
The function $\widehat{\varphi}$ satisfies the following:

\noindent(i)  Convexity: $\widehat{\varphi}(\mathbf x|\mathbf x^{[n]} )$ is convex in $\mathbf{x}$;

\noindent(ii) Upper bound: 
$\widehat{\varphi}(\mathbf x|\mathbf x^{[n]} )\geq \varphi(\mathbf{x})$;

\noindent(iii) Local equivalence: 
\begin{align}
    \widehat{\varphi}(\mathbf x^{[n]}|\mathbf x^{[n]} )= \varphi(\mathbf{x}^{[n]}),\ \nabla_{\mathbf{x}}\widehat{\varphi}(\mathbf x^{[n]}|\mathbf x^{[n]} )= \nabla_{\mathbf{x}}\varphi(\mathbf{x}^{[n]}).
\end{align}
\end{proposition}

Part (i) is proved by checking the semi-definiteness of the Hessian of $\widehat{\varphi}$.
Part (ii) is proved by checking 
$
\widehat{\varphi}(\mathbf x|\mathbf x^{[n]} )-\varphi(\mathbf x)
=
\frac{1}{\beta}\sum_{t=1}^T(x_{k}-x_{k}^{[n]})^2\geq 0$. 
Part (iii) is proved by comparing the function and gradient values of $\widehat{\varphi}$ and $\varphi$.
Based on part (i) of \textbf{Proposition 1}, problem $\text{P}2[n+1]$ is convex and can be solved by off-the-shelf software packages (e.g., Mosek).
Denote its optimal solution as
$\mathbf{x}^*$. 
Then, we set
$\mathbf{x}^{[n+1]}=\mathbf{x}^*$, such that the process repeats with solving the problem $\text{P}2[n+2]$.
According to parts (ii)--(iii) of \textbf{Proposition 1} and \cite[Theorem 1]{{sun2016majorization}}, every limit point of the sequence
$(\mathbf{x}^{[0]},\mathbf{p}^{[0]}),(\mathbf{x}^{[1]},\mathbf{p}^{[1]}),\cdots$ is a \emph{critical point} to problem $\text{P}2$ as long as the starting point $(\mathbf{x}^{[0]},\mathbf{p}^{[0]})$ is feasible to problem $\text{P}2$. 
The complexity of solving problem $\text{P}2$ (thus $\text{P}1$) with PMM is $\mathcal{O}(\mathcal{M}(2K)^{3.5})$, where $\mathcal{M}$ is the number of iterations for PMM to converge.

\vspace{-0.1in}
\section{Imitation Learning Optimization}
\vspace{-0.1in}

In Section \ref{sec:pmm}, we have developed a PMM method to solve the problem. However, the complexity of the PMM algorithm is high when $K$ is large, resulting in additional latency in problem solving. 

\begin{figure}[!t]
    \centering
    \includegraphics[width=0.49\textwidth]{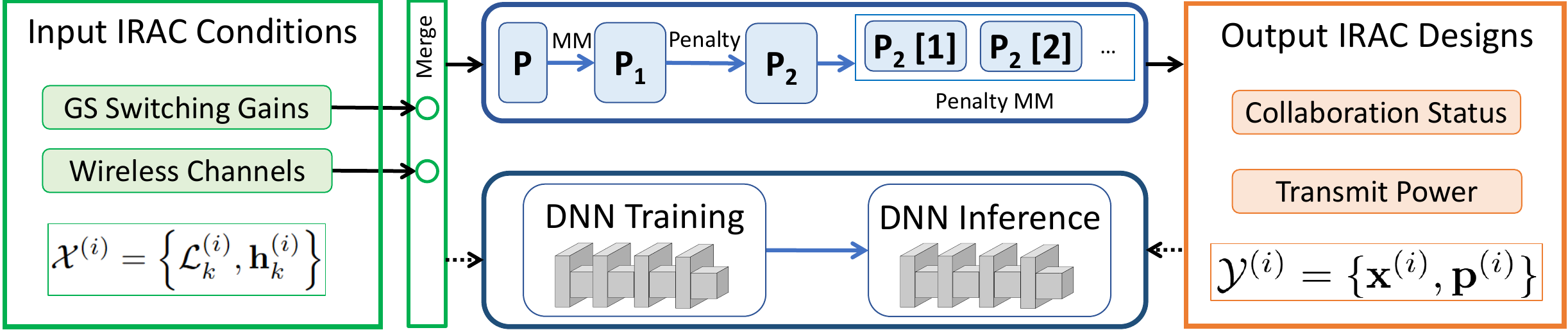} 
    \vspace{-0.25in}
    \caption{Pipeline of the ILO-IRAC algorithm.}
    \label{fig2} 
    \vspace{-0.25in}
\end{figure}

To address this issue, we propose an ILO method for faster IRAC design. 
The ILO trains an agent to solve $\text{P}1$ by imitating the PMM algorithm, such that time-consuming iterative computations are converted into real-time feed-forward inference \cite{shlezinger2023model,Liu2024survey}. 
The system architecture of ILO is shown in Fig. \ref{fig2}, which consists of three phases.

\textbf{1) Offline Demonstration Data Generation}: We adopt the PMM algorithm to generate the demonstration dataset 
$\mathcal{D} = \{\mathcal{S}^{(i)} \}_{i=1}^I$, where the $i$-th sample is given by $\mathcal{S}_i  = \{\mathcal{X}^{(i)},\mathcal{Y}^{(i)}\}$, and the input tuple is given by
$$\mathcal{X}^{(i)}=\left\{
\underbrace{\mathcal{L}(\Phi_\mathrm{edge}(\mathbf{s}_{k}^{(i)}),\Phi_{k}(\mathbf{s}_{k}^{(i)}))}_{:=\mathcal{L}_k^{(i)}},
\mathbf{h}_k^{(i)}\right\},$$
with $\mathcal{L}_k^{(i)}$ generated by GS model comparisons at sampled poses $\{\mathbf{s}_{k}^{(i)}\}$. 
Note that we can also include other factors $\{B_k,V_k,\sigma_k^2,T,P,S\}$ into $\mathcal{X}^{(i)}$. The output tuple 
$\mathcal{Y}^{(i)}=\{\mathbf{x}^{(i)},\mathbf{p}^{(i)}\}$ is generated by executing the PMM algorithm on problem configuration $\mathcal{X}^{(i)}$. 
    
\textbf{2) Offline DNN Training}: The DNN consists of three fully-connected layers with $100$, $72$, and $20$ units, respectively. The first two hidden layers utilize the rectified linear unit (ReLU) activation function. For model training, we use the Focal Loss function and optimize the network parameters using the AdamW algorithm.

\textbf{3) Online DNN Inference}: The pre-trained DNN model is deployed at the edge server to make real-time decisions. At each time step $t$, the server estimates the current rendering and communication conditions $\mathcal{X}^{(t)}$ using users' poses and pilots.
Feeding $\mathcal{X}^{(t)}$ into the DNN model, we can obtain the optimal collaboration status $\mathbf{x}^{(t)}$ and transmit power $\{\mathbf{p}^{(t)}\}$.

\vspace{-0.1in}
\section{Experiments}
\vspace{-0.1in}

We implement the proposed ECO-GS system with PMM and ILO algorithms in Python based on the 3D GS project \cite{kerbl20233d}. We deploy the system on a Linux workstation with AMD 3.7\,GHz CPU and NVIDIA A6000 GPU.
We conduct experiments exploiting the truck dataset \cite{kerbl20233d}.

We consider the case of $N=600$ and $(K,S)=(20,10)$ \cite{wang2020angle}.  
Edge-user distances $\{d_k\}$ are computed based on their locations, where users are randomly located within a $100 \times 100\,\text{m}^2$ area and edge server is at $(0, 0)$. 
Channels are generated by $\mathbf{h}_k\sim\mathcal{CN}(0,\varrho_k\mathbf{I}_N)$, where $\mathcal{CN}$ is complex Normal distribution, $\mathbf{I}_N$ is $N\times N$ identity matrix, and pathloss $\varrho_k=d_k^{-\alpha}$ with $\alpha = 3$ \cite{wang2020angle}. 
The noise power $\sigma^2 = -70 \,\text{dBm}$ and the bandwidth is $B_k=2\,\text{MHz}$.
The end-to-end time budget is $T=60$\,ms, the edge rendering time is $T_0=6.5$\,ms (tested 
$\Phi_{\mathrm{edge}}(\cdot)$ on A6000 server), and the user rendering time is $16.7$\,ms (tested $\Phi_{k}$ on a desktop computer).
All quantitative results are obtained by averaging over $100$ random simulation runs, with independent channels and user views in each run.

We compare ECO-GS with PMM and ILO with the following baselines:
1) \textbf{UserGS}: GS \cite{kerbl20233d} (but with model compression) (i.e., $\{x_k=0\}$);
2) \textbf{MaxRate}: ECO-GS with sum-rate maximization \cite{zhang2024efficient};
3) \textbf{Greedy}: ECO-GS with large-$\mathcal{L}_k$-first principle \cite{2025TCCN};
4) \textbf{Search}: ECO-GS by solving $\mathsf{P}_{\mathrm{GS}}$ with iterative local search \cite{neumann2007randomized};
5) \textbf{Rounding}: ECO-GS by solving $\mathsf{P}_{\mathrm{GS}}$ with continuous relaxation and rounding \cite{hubner2014rounding};

 \begin{table}[!t]
\caption{Large GS Versus Small GS}
\vspace{-0.1in}
\label{tab:GS}
\centering
\scalebox{0.7}{
\begin{tabular}{cccccccc}
    \toprule
    Model & Platform & $\downarrow$Dim. & $\downarrow$Model Size & {$\uparrow$PSNR} & {$\uparrow$SSIM} 
    & {$\downarrow$L1}
    & {$\downarrow$Loss} \\
    \midrule
    $\Phi_{\text{edge}}$  & Edge  & 64 & 509.8\,M  & 27.49 & 0.98 & 0.031 & 0.029 \\
    \rowcolor{blue!10} $\Phi_{k}$     & Device & 14 & 9.3\,M & 24.99 & 0.97 & 0.043 & 0.041 \\
    \bottomrule
    \vspace{-0.15in}
\end{tabular}
}
\vspace{-0.2in}
\end{table}

\begin{figure}[!t]
	\centering
	\begin{subfigure}{0.48\linewidth}
		\centering
		\includegraphics[width=1\linewidth]{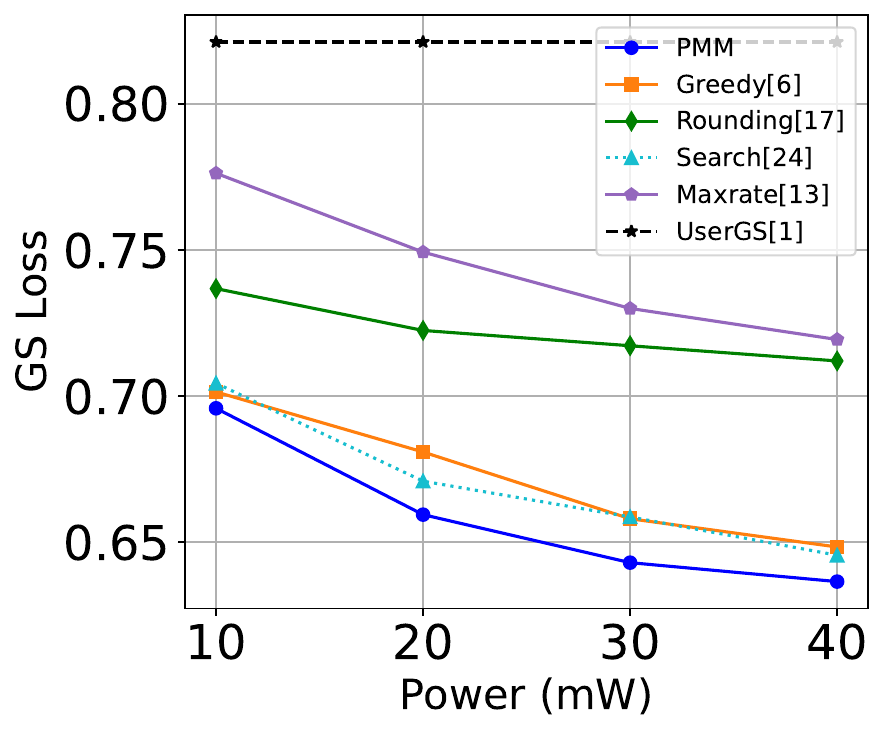}
		\vspace{-0.25in}
        \caption{GS Loss versus $P$.}
	\end{subfigure}
 	\begin{subfigure}{0.48\linewidth}
		\centering
		\includegraphics[width=1\linewidth]{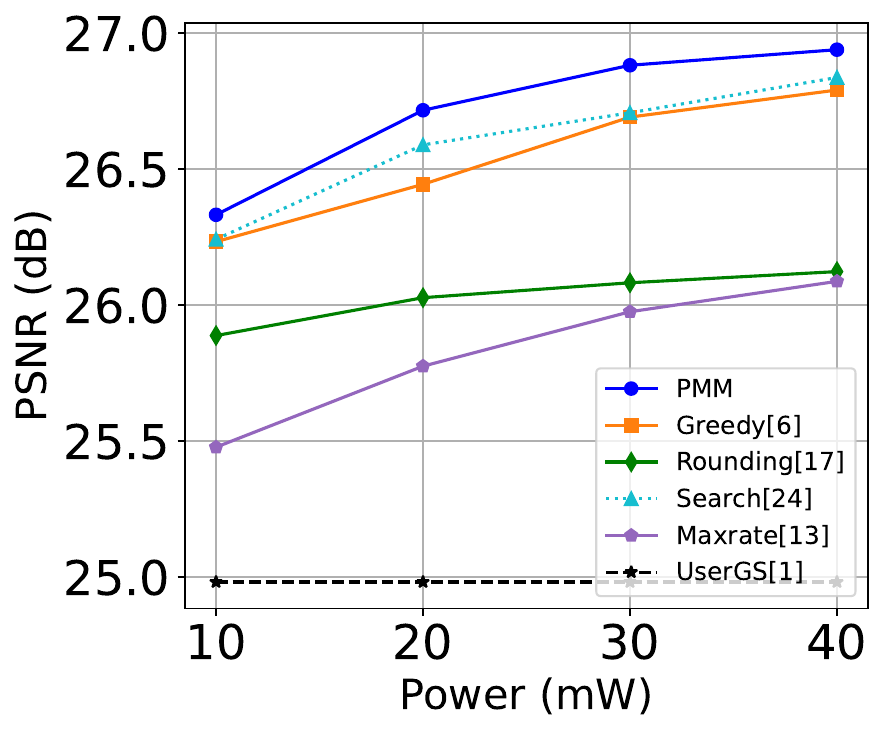}
		\vspace{-0.25in}
        \caption{PSNR versus $P$.}
	\end{subfigure}
    \vspace{-0.1in}
	\caption{Quantitative comparison of different schemes.}
    \vspace{-0.1in}
	\label{compare}
\end{figure}

\begin{figure}[t]
    \centering
    \includegraphics[width=0.48\textwidth]{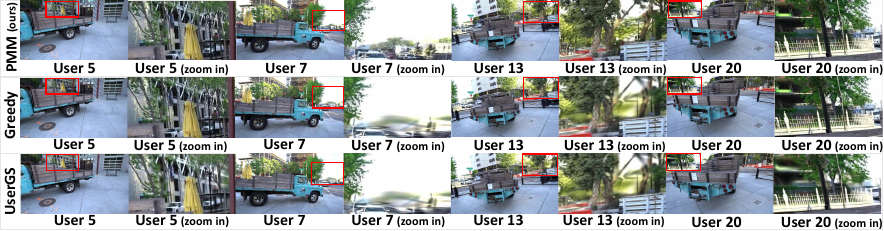}
    \vspace{-0.2in}
    \caption{Comparison of rendered images.}
    \vspace{-0.15in}
    \label{fig:render}
\end{figure}

Table I compares small and large GS models. 
The small GS model is obtained by reducing the dimension of each Gaussian from $64$ to $14$.
It reduces the model size by \textbf{over $50$x} compared to the large GS model. 
This demonstrates the effectiveness of employing model compression for low-cost GS deployment. 
However, the PSNR and SSIM performances are degraded, meaning that remote rendering is necessary if small GS models fail in rendering qualified images.

Next, we compare the proposed IRAC with PMM to other benchmark schemes under $P=\{10,20,30,40\}$\,mW.
It can be seen from Fig.~\ref{compare} that the proposed IRAC with PMM consistently outperforms all the other schemes. 
Compared with UserGS, the loss reduction is at least $15\%$, and the improvement of PSNR is over $1.3$\,dB.

\begin{figure}[!t]
	\centering
        \begin{subfigure}{0.45\linewidth}
		\centering
		\includegraphics[width=\linewidth]{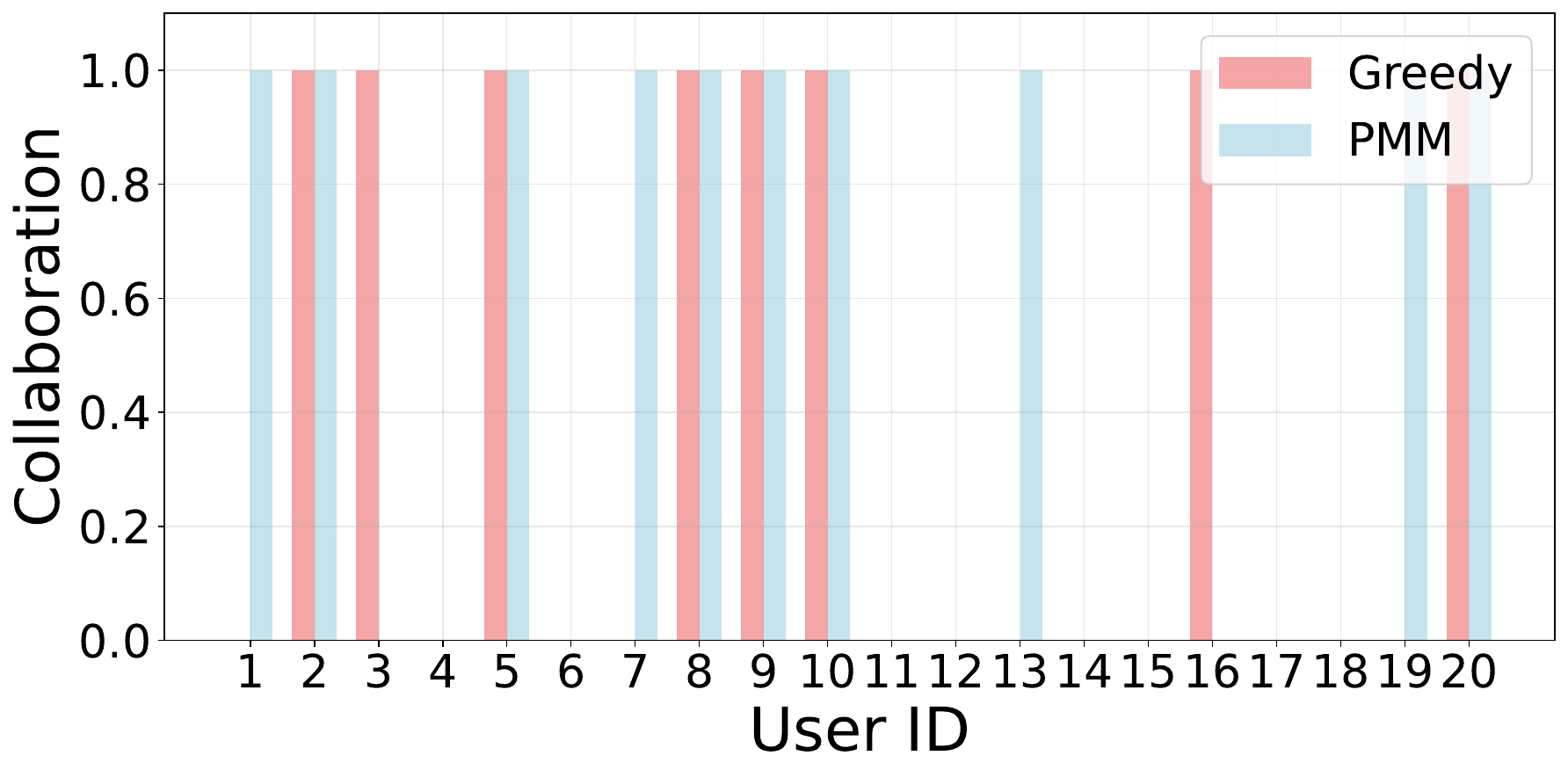}
      \vspace{-0.2in}
		\caption{Collaboration statuses.}
	\end{subfigure}
    	\centering
	\begin{subfigure}{0.45\linewidth}
		\centering
		\includegraphics[width=\linewidth]{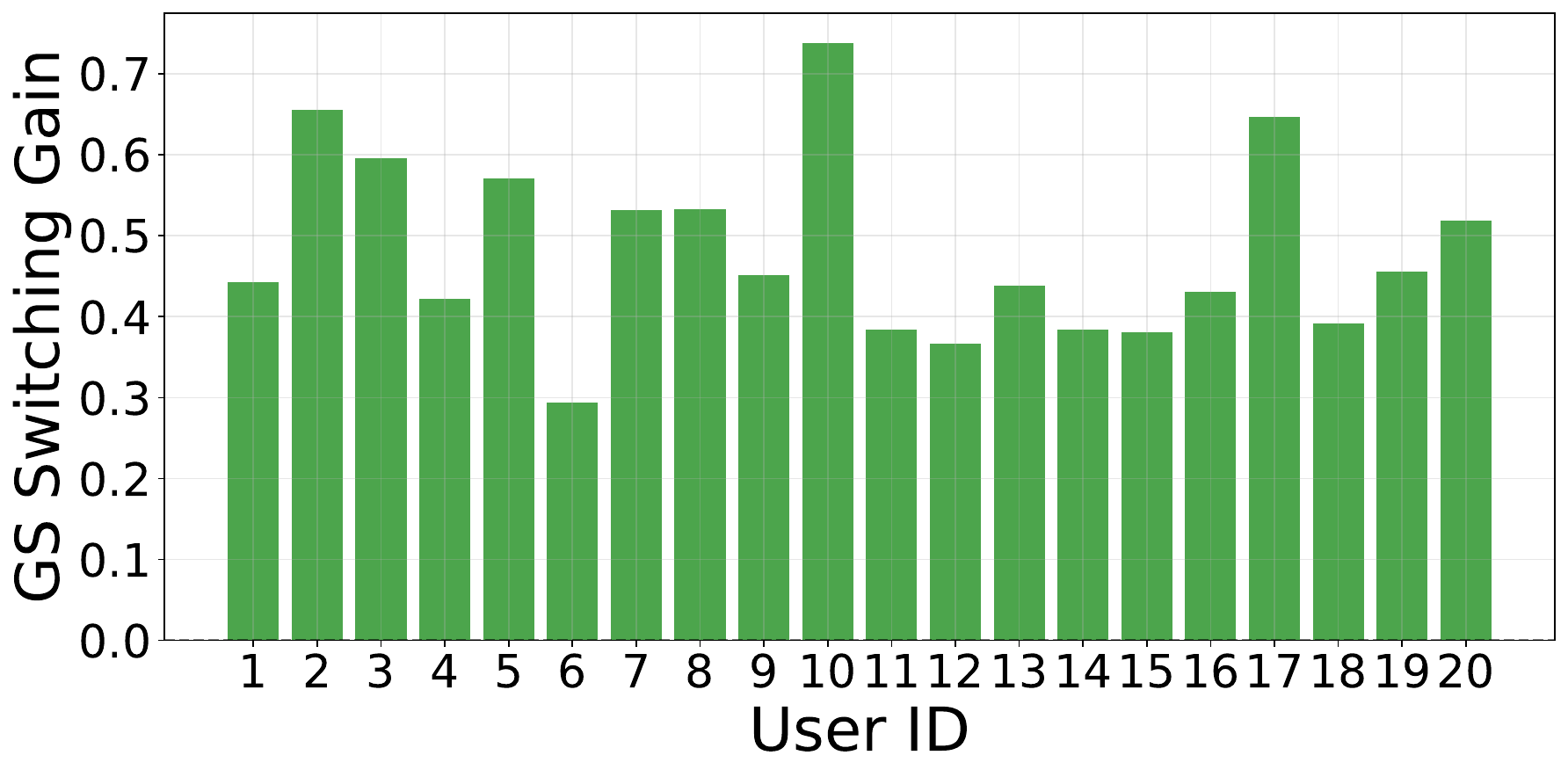}
      \vspace{-0.2in}
		\caption{GS switching gains.}
	\end{subfigure}
	\begin{subfigure}{0.45\linewidth}
		\centering
		\includegraphics[width=\linewidth]{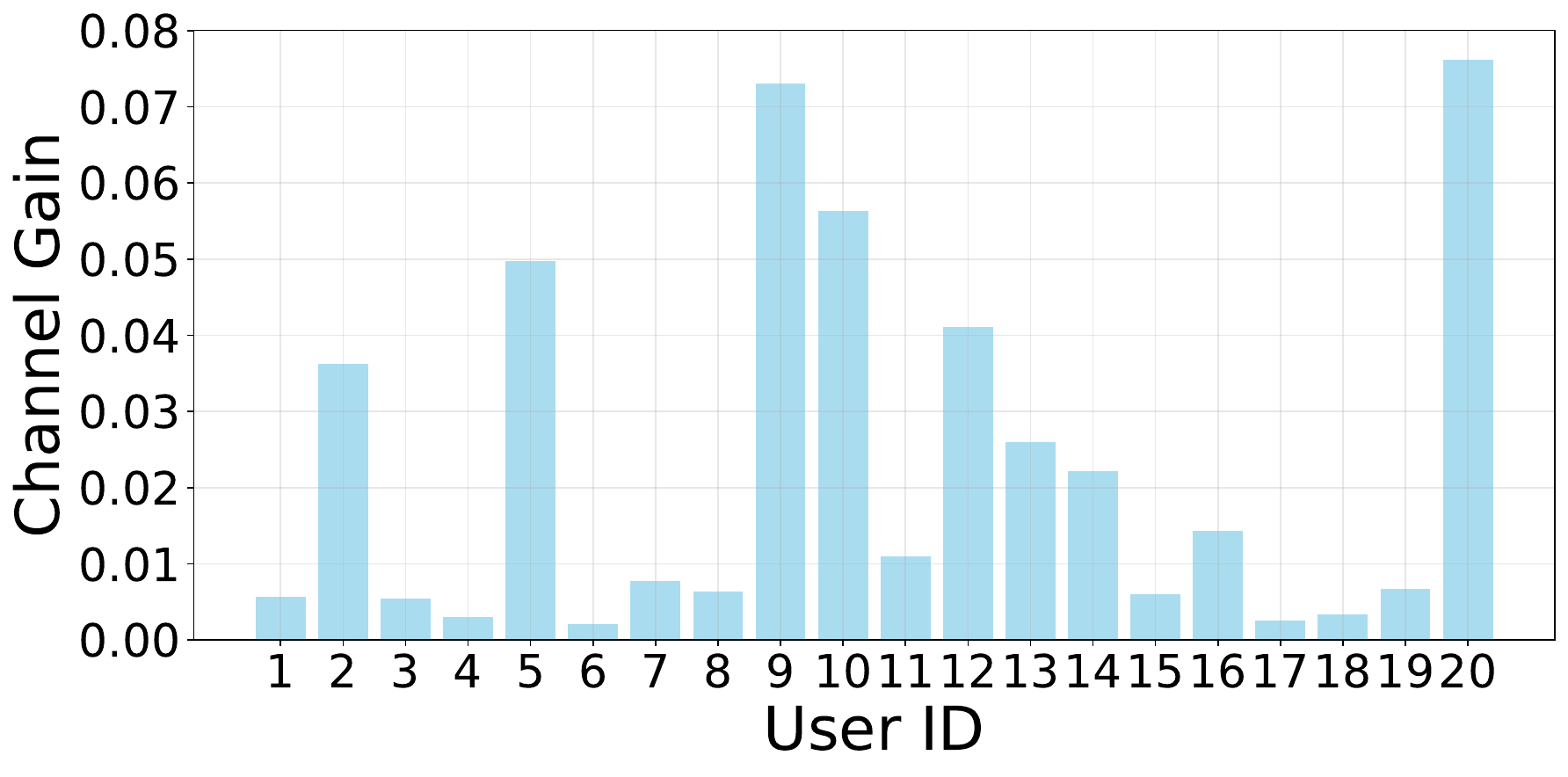}
        \vspace{-0.2in}
		\caption{Channel gains.}
	\end{subfigure}
    \begin{subfigure}{0.45\linewidth}
		\centering
		\includegraphics[width=\linewidth]{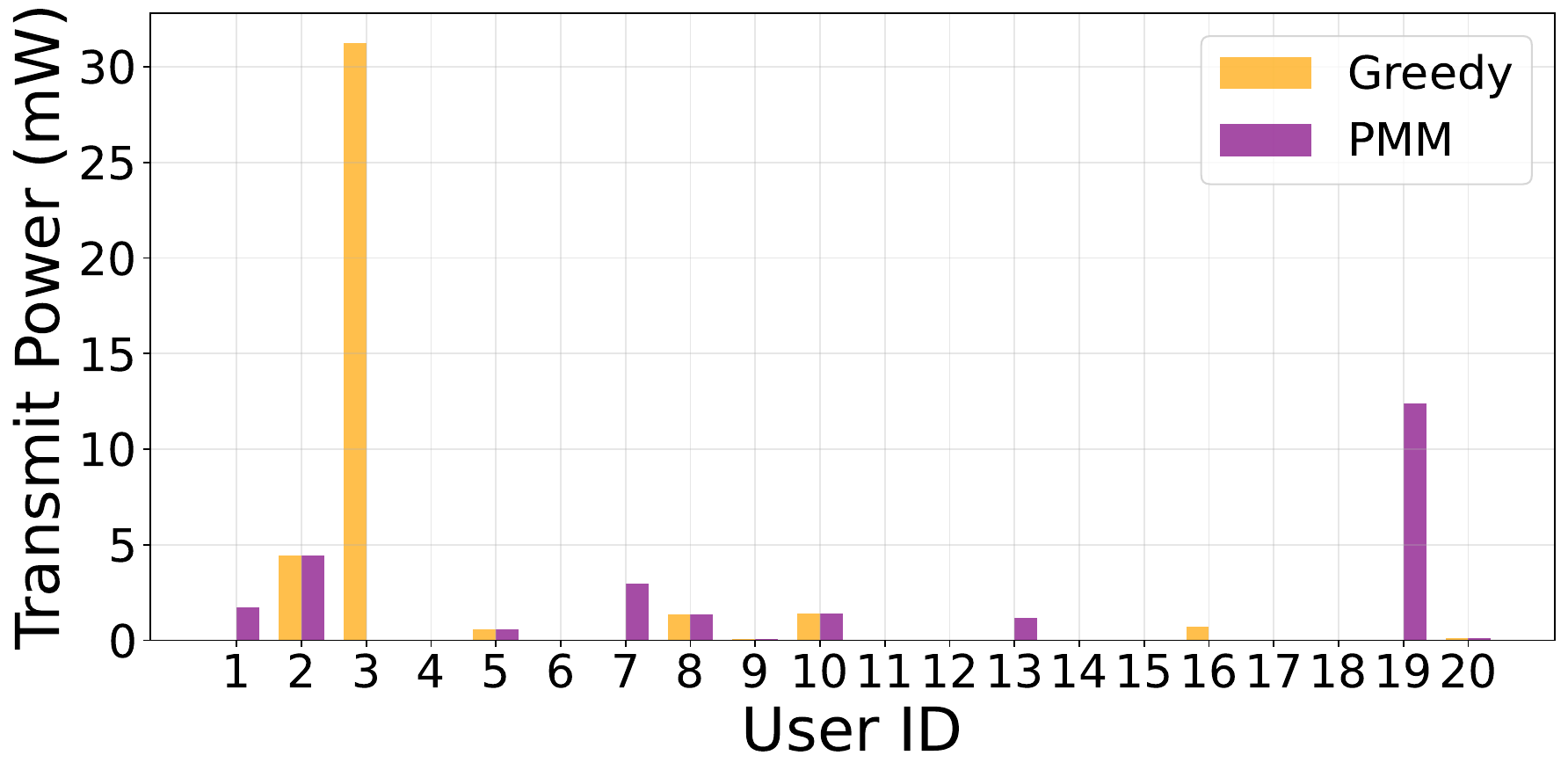}
      \vspace{-0.2in}
		\caption{Transmit powers.}
	\end{subfigure}
     \vspace{-0.1in}
	\caption{
    Case study for PMM and Greedy schemes.}
     \label{fig:demo}
     \vspace{-0.15in}
\end{figure}

To obtain deeper insights, we consider the case of $P_{\text{sum}}=40$\,mW, and the rendering results are shown in Fig. \ref{fig:render}. 
It can be seen that all the images of PMM have high qualities.
This corroborates Fig. \ref{fig:demo}a and Fig. \ref{fig:demo}b, where the selected users have high GS switching gains.
This implies that PMM is able to distinguish heterogeneous rendering requirements. 
We also observe that images $(7,13)$ are blurred for the Greedy scheme.
This is because the Greedy scheme selects user $3$ with higher GS loss for collaboration. However, this user experiencing bad channel conditions costs excessive power resources.
In contrast, our scheme avoids collaborating remote users as shown in Fig. \ref{fig:demo}c and Fig. \ref{fig:demo}d.
Lastly, all images are blurred for the UserGS scheme, which demonstrates the necessity of leveraging edge GS to compliment local GS.

\begin{figure}
		\centering
		\includegraphics[width=0.8\linewidth]{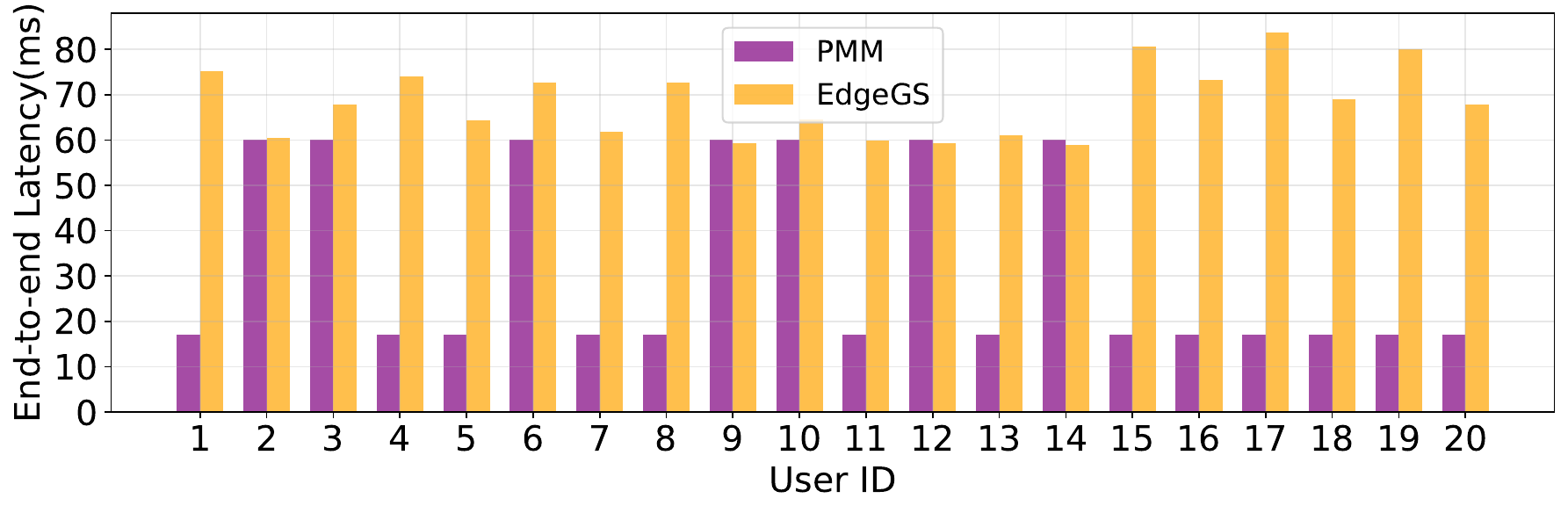}
      \vspace{-0.1in}
		\caption{Case study for PMM and EdgeGS schemes.}
        \vspace{-0.05in}
        \label{demo2}
\end{figure}

One may wonder whether we can adopt remote rendering for all users, i.e., $\{x_k=1\}$. This is known as the \textbf{EdgeGS} scheme \cite{Mehrabi2021MultiTier}, which may lead to non-smooth rendering due to latency. 
To see this, the end-to-end latency at $P_{\text{sum}}=10$\,mW is provided in Fig. \ref{demo2}. It can be seen that EdgeGS leads to over $80$\,ms latency, while our method strictly guarantees less than $60$\,ms for smooth rendering.

\begin{figure}[!t]
	\centering
	\begin{subfigure}{0.45\linewidth}
		\centering
		\includegraphics[width=1\linewidth]{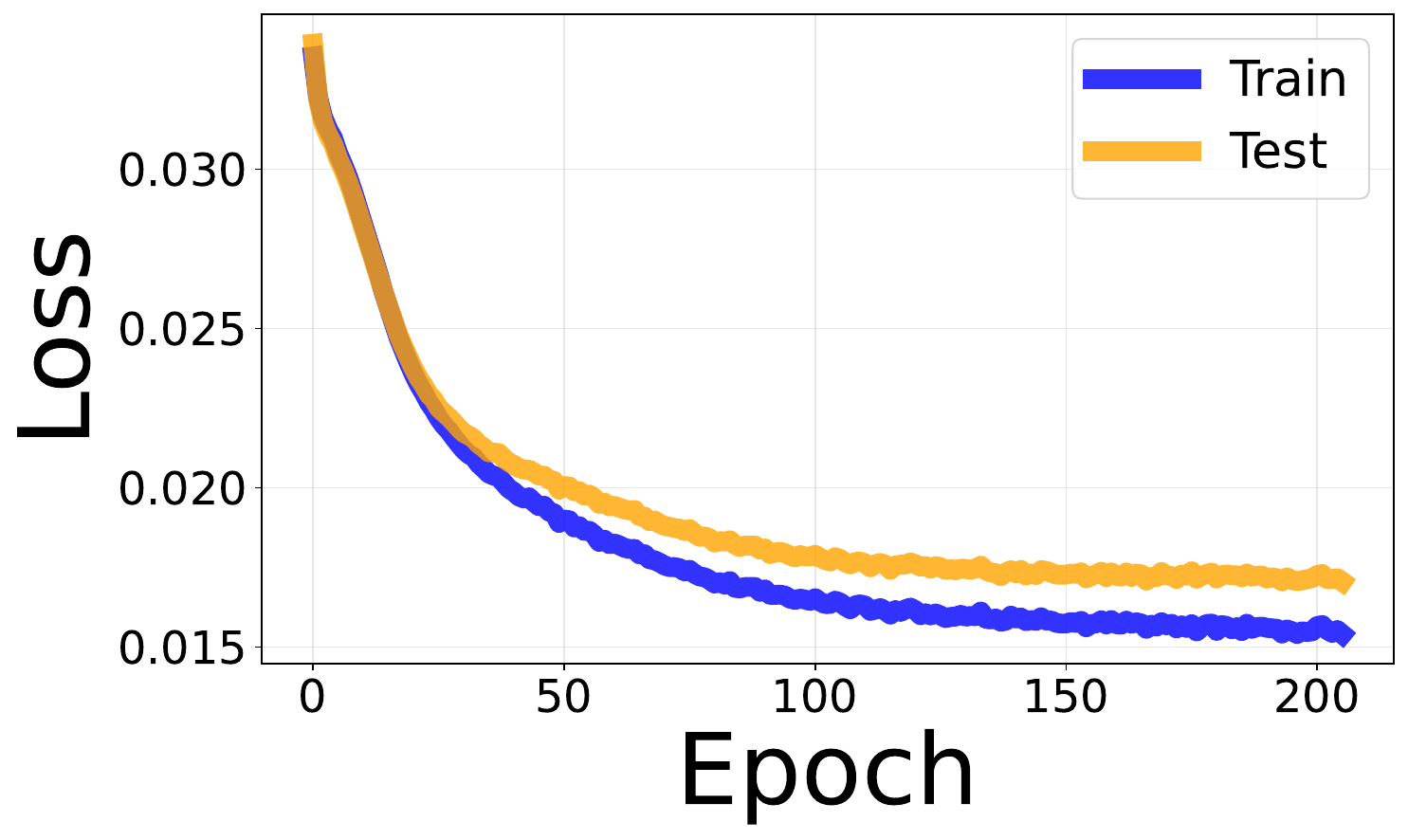}
             \vspace{-0.2in}
		\caption{Loss versus Epoch.}
	\end{subfigure}
 	\begin{subfigure}{0.45\linewidth}
		\centering
		\includegraphics[width=1\linewidth]{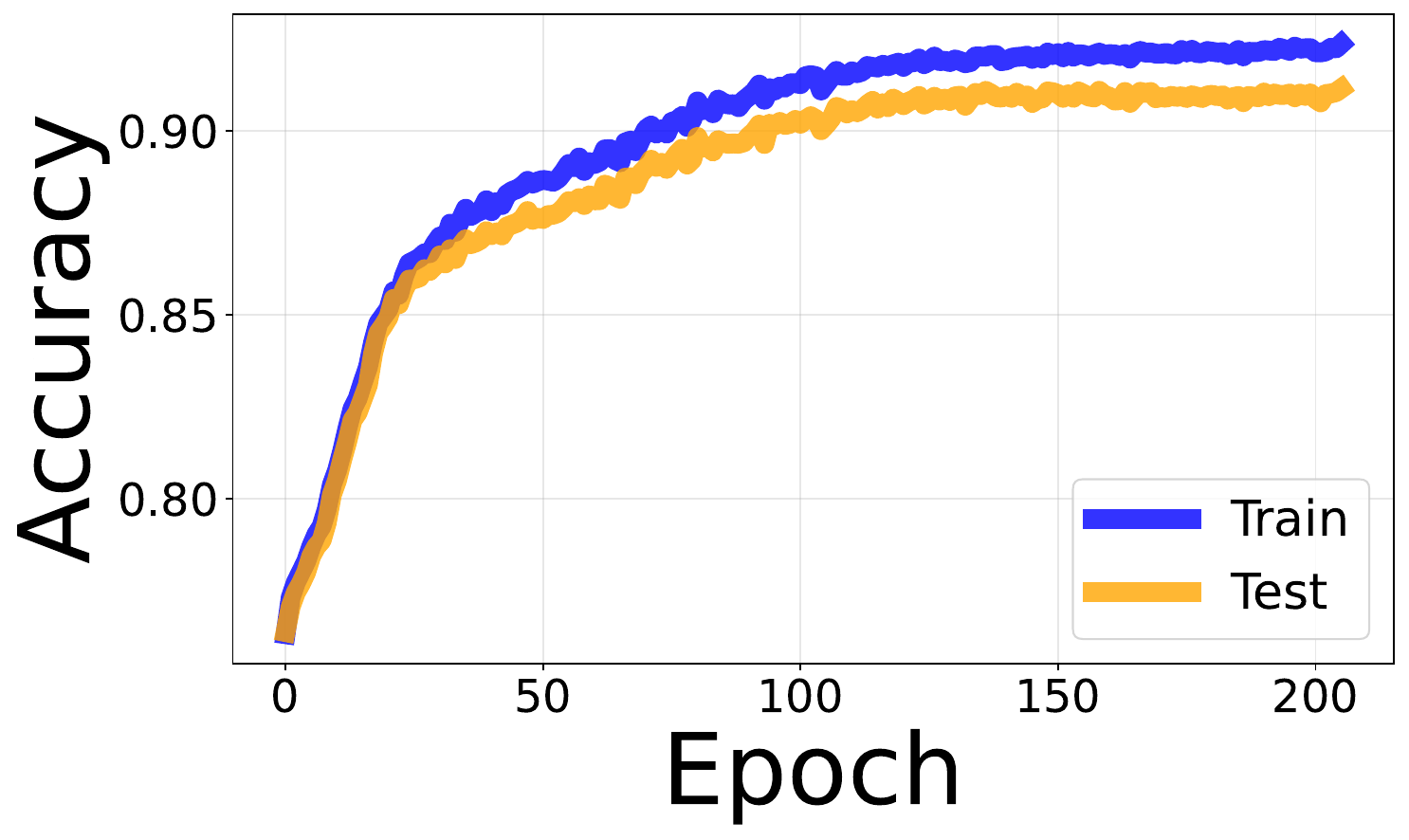}
             \vspace{-0.2in}
		\caption{Accuracy versus Epoch.}
	\end{subfigure}
     	\begin{subfigure}{0.45\linewidth}
		\centering
		\includegraphics[width=1\linewidth]{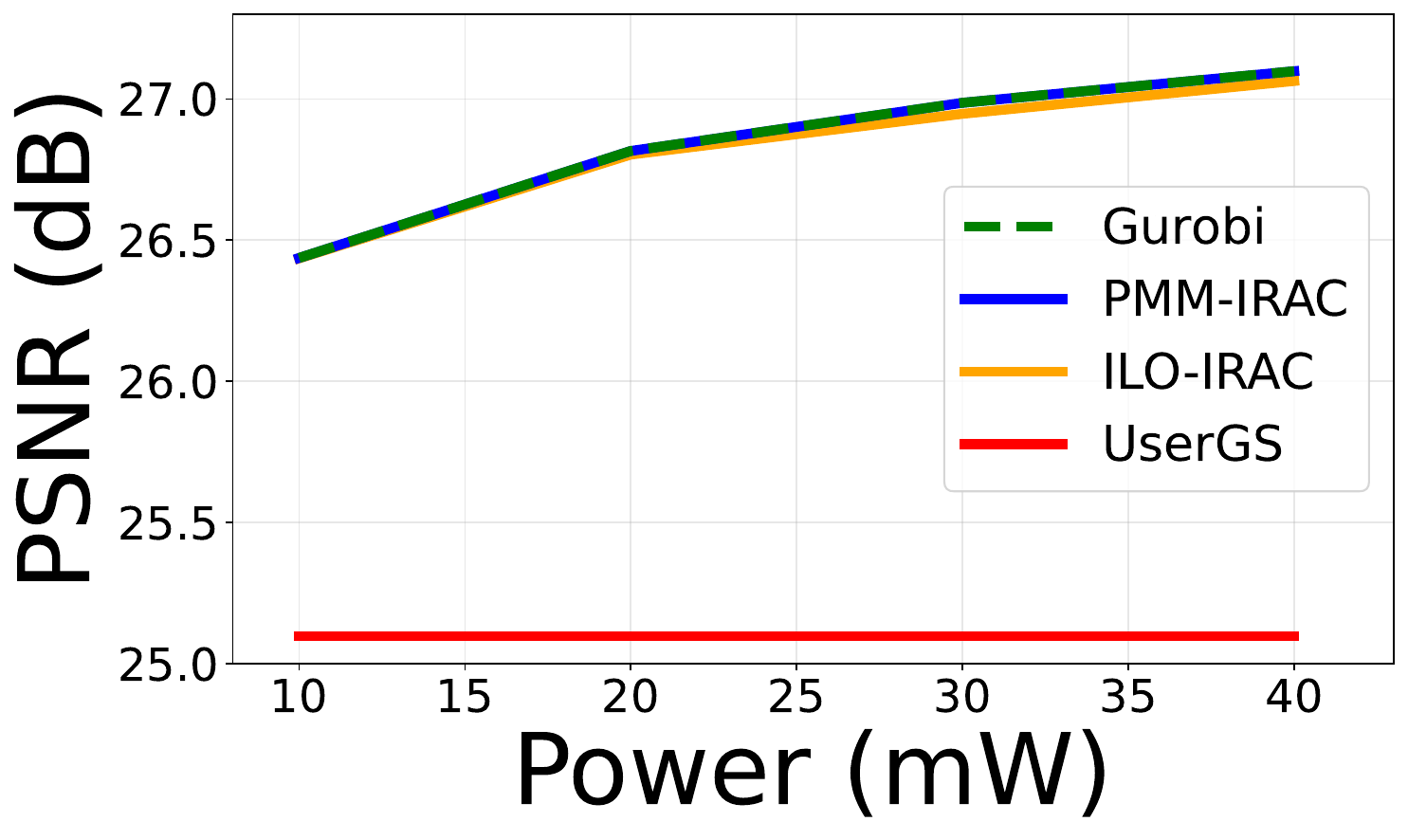}
             \vspace{-0.2in}
		\caption{PSNR versus $P$.}
	\end{subfigure}
         	\begin{subfigure}{0.45\linewidth}
		\centering
		\includegraphics[width=1\linewidth]{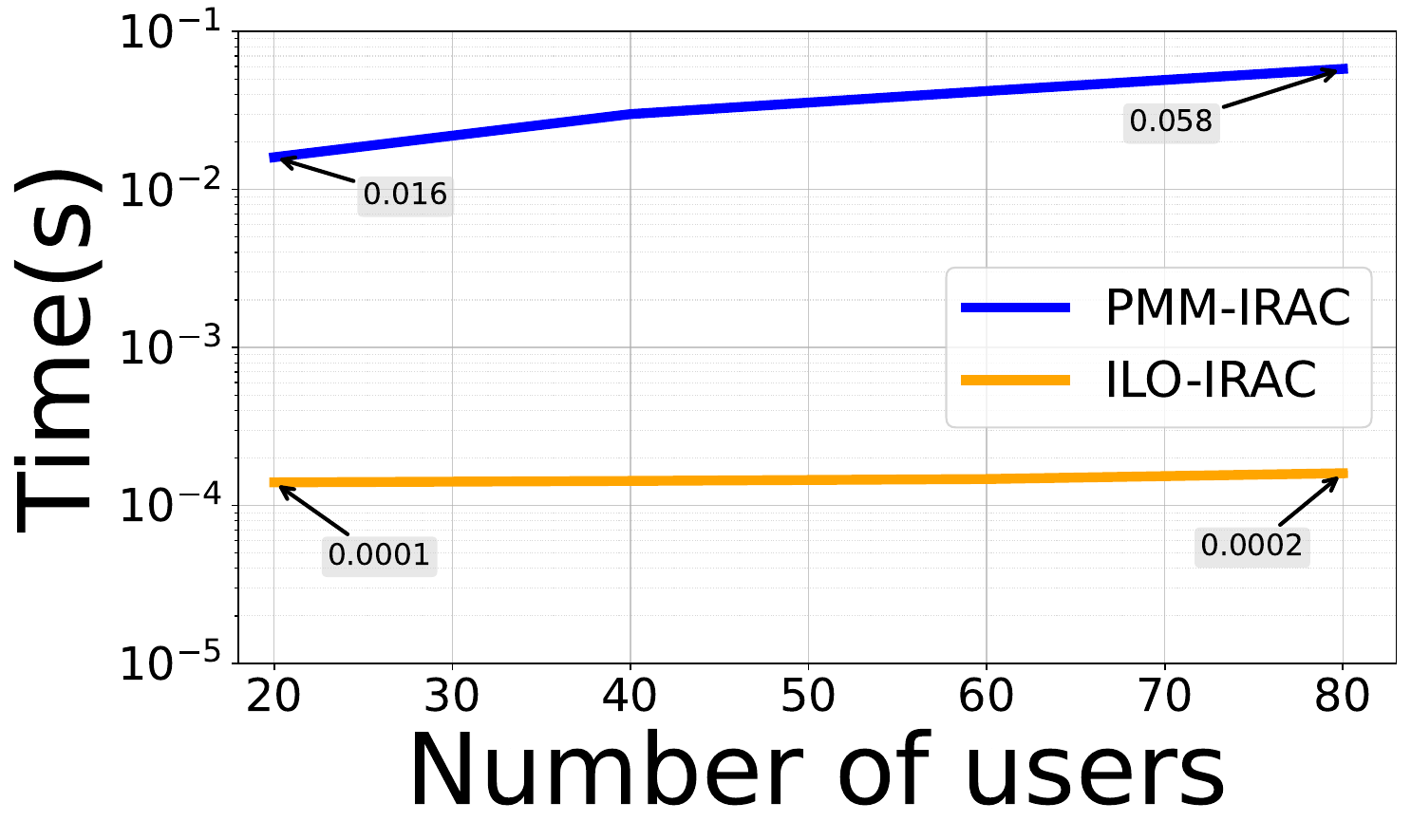}
             \vspace{-0.2in}
		\caption{Execution Time versus $K$.}
	\end{subfigure}
        \vspace{-0.1in}
	\caption{Evaluation of ILO-IRAC.}
	\label{fig:ILO}
    \vspace{-0.25in}
\end{figure}

Finally, we evaluate the proposed ILO-IRAC. 
We generate $10000$ random samples of $\{\mathcal{X}^{(i)},\mathcal{Y}^{(i)}\}$, including $5000$ training and $5000$ test samples.
We train the ILO DNN for $200$ epochs with a learning rate of $6\times 10^{-4}$ and batch size of $96$. 
Fig.~\ref{fig:ILO}a and Fig.~\ref{fig:ILO}b illustrate the training and testing losses and accuracies, respectively.
It can be seen that the proposed ILO-IRAC achieves an accuracy of $0.9$ and a loss of $0.0175$, indicating that ILO can learn how to imitate PMM. 
The PSNR performances and computation times of ILO-IRAC and PMM-IRAC are compared in Fig.~\ref{fig:ILO}c and Fig.~\ref{fig:ILO}d.
ILO reduces the time by $100$x compared to that of PMM, while achieving a competitive PSNR with only $0.02$\,dB degradation. 
The computation time of ILO-IRAC is within $1$\,ms, making it suitable for real-time execution.

\vspace{-0.1in}
\section{Conclusion}\label{section7}
\vspace{-0.1in}

This paper has presented the IRAC framework for ECO-GS, which mitigates the discrepancies between GS-rendered images and real-world environments under communication constraints.
We have proposed two algorithms to obtain the best trade-off between timeliness and fidelity in IRAC designs. 
Extensive experiments have demonstrated that collaborative rendering indeed helps, but we need to distinguish non-beneficial users. 
Furthermore, remote users should be carefully treated, as they may cost excessive transmit powers.

\clearpage
\newpage

\vfill\pagebreak

\bibliographystyle{IEEEbib}

\end{document}